# Neutronics Calculation Advances at Los Alamos: Manhattan Project to Monte Carlo

Avneet Sood*, R. Arthur Forster, B.J. Archer, and R.C. Little
*Los Alamos National Laboratory, Los Alamos, NM 87545*
*Los Alamos National Laboratory, P.O. Box 1663, MS F663
*E-mail: sooda@lanl.gov

**Abstract** - The history and advances of neutronics calculations at Los Alamos during the Manhattan Project through the present is reviewed. We briefly summarize early simpler, and more approximate neutronics methods. We then motivate the need to better predict neutronics behavior through consideration of theoretical equations, models and algorithms, experimental measurements, and available computing capabilities and their limitations. These, coupled with increasing post-war defense needs, and the invention of electronic computing led to the creation of Monte Carlo neutronics transport. As a part of the history, we note the crucial role that the scientific comradery between the great Los Alamos scientists played in the process. We focus heavily on these early developments and the subsequent successes of Monte Carlo and its applications to problems of national defense at Los Alamos. We cover the early methods, algorithms, and computers, electronic and women pioneers, that enabled Monte Carlo to spread to all areas of science.

Keywords: Manhattan Project, Los Alamos National Laboratory, von Neumann, Ulam, Metropolis, Fermi, Richtmyer, ENIAC, MANIAC, neutronics, Monte Carlo, MCNP

## I. THE STATE OF NEUTRONICS CALCULATIONS DURING WORLD WAR II

### *I.A. Discovery of Fission*

In December 1938, physicists Lise Meitner and Otto Frisch made a revolutionary discovery that would forever change nuclear physics and lead the world into the atomic age: a uranium nucleus had fissioned. The prior discovery of the neutron in 1932 had created the capability for scientists to probe the atomic nucleus. In 1934, scientists like Enrico Fermi bombarded large, heavy nuclei with neutrons to produce the first elements heavier than uranium—or so he thought. The thinking was that hitting a large nucleus with a neutron could only induce small changes in the number of neutrons or protons. Other scientist teams like Lise Meitner and her nephew, Otto Frisch, Austrians exiled in Sweden and Otto Hahn and Fritz Strassmann in Berlin also began bombarding uranium and other heavy elements with neutrons and found isotopes of lighter nuclides among the decay products. Meitner and Frisch, using the liquid-drop model of the nucleus, were able to estimate that approximately 200 MeV would be required to split a uranium nucleus. Meitner was able to measure the daughter nuclei's mass, and show that it was less by about 200 MeV. Meitner and Frisch sent their paper to Nature in January. Frisch named the process "fission" after observing biologists describing cell division as "binary fission". Hahn and Strassmann published their findings separately. Bohr carried the news of the discovery of fission to America. Other scientists discovered that the fission reaction also emitted enough secondary neutrons for a chain reaction along with the release of an enormous amount of energy. As word spread, the worldwide scientific community realized the possibility of both making weapons and getting power from fission. The MAUD report,[1] a secret report by British scientific leaders, warned of the likelihood of producing a weapon if the U-235 isotope could be separated. The creation of such atomic weapons would have a massive impact on the war. The MAUD report estimated that a critical mass of 10 kg would be sufficient for a weapon, small enough to load onto an airplane, and ready in approximately two years. The MAUD report along with the US Government's National Academy of Sciences report helped launch the Manhattan Project and the start of Project Y, Los Alamos. The race for a weapon was on.[2,3]

### *I.B. Early Understanding of Fission*

Enrico Fermi, Frederic Joliot, Leo Szilard, and others independently found secondary neutrons from fission experiments that they conducted in Paris and in New York. Weapons applications for fission required the geometric progression of a chain reaction of fission events. [Serber notes that the term "to fish" was used in place of "to fission" but the phrase did not stick]. These scientists had an early understanding of how neutron chain reactions could be used to release massive amounts of energy. The uncontrolled fissioning of U235 resulting in a massive amount of energy makes the material very hot, developing a great pressure and thus exploding. Basic equations characterizing the energy released had been developed but assumed an ideal geometric arrangement of materials where no neutrons are lost to the surface or to parasitic absorptions. In realistic configurations, neutrons would be lost by diffusion outward through the surface. The precise dimensions for a perfect sphere needed to be determined to exactly balance between the chain reaction of neutrons created by fission and the surface and parasitic losses. This critical radius would also depend on the density of the material. An effective explosion depends on whether this balance of losses occurred before an appreciable fraction of the material had fissioned releasing enormous amounts of energy in a very short time. Slow, or thermal energy, neutrons cannot play an effective role in an explosion since they require significant time to be captured to cause fission.

Hans Bethe later described assembling fissionable material together and the resulting complicated shapes. Bethe describes that these problems were insoluble by analytical means and the resulting equations were beyond the existing desk computing machines.[4]

It wasn't until early 1945 that Los Alamos received amounts of uranium and plutonium with masses comparable to their critical mass so that experiments could be performed to compare with the early predictions. Plutonium and uranium were fabricated into spheres and shells for comparison with calculations. The critical masses were determined experimentally by a series of integral experiments. Early critical experiments were done using reflected geometries.[5,6] The critical radius was determined by plotting the inverse of the observed neutron multiplication as increasingly more material was added and extrapolating for the point of infinite multiplication. This paper discusses the various computational methods for solving the neutron transport equation during the Manhattan Project, the evolution of a new computational capability, and the invention of stochastic methods, focusing on Monte Carlo, and nuclear data.

### *I.C. Manhattan Project Advances Solutions to the Neutron Transport Equation*

The earliest estimates for a nuclear bomb were based on natural uranium, and gave critical masses of many tons. In March 1940 Rudolf Peierls and Otto Frisch used diffusion theory to estimate that the critical mass of U235 was less than 1 kg.[7] This estimate persuaded both the United Kingdom and United States to proceed with a nuclear bomb program. Knowing the critical mass of U235 and Pu239 were essential for the Manhattan Project for sizing the production plants, and setting safety limits on experiments. This requirement drove the Los Alamos Theoretical Division (T-Division) to constantly improve the methods to determine the critical masses.

*I.C.1. Neutron Transport Equation*

Estimating the critical masses required solutions to the neutron transport equation. The neutron transport equation is a mathematical description of neutron multiplication and movement through materials and is based on the Boltzmann equation which was used to describe study the kinetic theory of gasses. The Boltzmann equation was developed in 1872 by Ludwig Boltzmann to explain the properties of dilute gasses by characterizing the collision processes between pairs of molecules.[8] It is a mathematical description of how systems that are not in a state of equilibrium evolve (e.g., heat flowing from hot to cold, fluid movement etc.) and can be used to describe the macroscopic changes in a system. It has terms describing collisions and free streaming. The general equation looks like:

$$\frac{\partial f}{\partial t} + v \cdot \nabla_r f + \frac{K}{m} \nabla_v f = \int d\Omega \int dv \, \sigma(\Omega)|v - v'| (f' - f)$$

f(r,v,t) is particle distribution and is a function of position (r), velocity (v), and time (t); F is an external force acting on the particles. The left-hand-side of the equation represents the free-streaming of particles including the time-rate of change, geometric losses, and gains from other sources. The right-hand-side of the equation represents the collisions occurring during the flow. The equation is a non-linear integro-differential equation involving terms that depends on particle position and momentum.[9] A linearized form of neutron transport equation can be cast into this general Boltzmann form to describe neutron multiplication and movement through materials.[10] The integro-differential form of the linear, time-dependent neutron transport equation can be written as:

$$\left(\frac{1}{v}\frac{\partial}{\partial t} + \Omega \cdot \nabla + \Sigma_t\right)\psi = \frac{\mathcal{X}_p}{4\pi}\int dE' \nu \Sigma_f \phi + \sum_{i=1}^{N} \frac{\mathcal{X}_{di}}{4\pi}\lambda_i C_i + \int d\Omega' \int dE' \Sigma_s \psi + Q$$

where $\psi = \psi(r, E, \Omega, t)$ is the angular neutron flux, $\phi = \phi(r, E, t)$ is the scalar neutron flux, $\mathcal{X}_p, \mathcal{X}_d$ are the fission and delayed neutron precursor exit energy distributions, $\Sigma_t, \Sigma_s, \Sigma_f$ are the macroscopic total, scattering, and fission cross sections and dependent on $r, E, \Omega, t$, and $\lambda_i C_i$ are the decay constant and total number of precursor, i, N is the total number of delayed neutron precursors, and $Q = Q(r, E, \Omega, t)$ source term. The neutron transport equation is a balance statement that conserves neutrons where each term represents a gain or a loss of neutrons. The neutron equation states that neutrons gained equal neutrons lost. The equation is usually solved to find the scalar neutron flux, $\phi$, thus allowing for the calculation of reaction rates. Reaction rates can be measured and are the primary interest in most applications.

Today this equation can be solved directly by either deterministic or stochastic methods using advanced algorithms and computational resources. Deterministic methods are exact solutions everywhere in a problem to an approximate and discretized equation. Monte Carlo methods are a statistically-approximate solution to specific regions of a problem to an exact equation. The techniques are complimentary and can be used separately or in tandem. The neutron transport equation can be simplified by making several assumptions allowing for analytic solutions or simple calculations that could be carried out by the electro-mechanical computers available early in Manhattan Project. Significant advances in solving simplified versions of the neutron transport equation were made during this time.

*I.C.2. Approximations and Limitations in Diffusion Theory*

Fick's law states that particles will diffuse from the region of higher flux to the region of lower flux through collisions. It is valid for locations far (several neutron mean free paths [mfp]) from the boundaries because of the initial infinite medium assumption. Additionally, since the neutron flux was assumed to be predominately due to isotropic scattering collisions, proximity to strong sources or absorbers or any strongly an-isotropic materials invalidates the original derivations. This invalidation also includes proximity to any strongly varying material interfaces or boundaries. Lastly, Fick's law assumed that the flux was independent of time. Rapid time-varying flux changes (the time for few mfp collisions) were shown to be the upper limit. Many of these approximations were not applicable for weapons work but could be useful for a general predictive capability. Limitations on the methods were understood and better methods were being developed. Notably, work by Eldred Nelson and Stan Frankel from the Berkeley theoretical group for a more exact diffusion theory-based solution dramatically improved the accuracy of solutions.[11]

*I.C.3. Analytic Solutions*

The first improvement in generating analytic solutions came in the late spring of 1942 when Oppenheimer and Serber tasked Stanley Frankel and Eldred Nelson to verify the British critical mass calculations.[3] It turned out that the differential form of the diffusion equations with realistic boundary conditions was difficult to solve analytically. At the time, a very general integral equation was being used in broader science applications to describe physical phenomena and was being mathematically solved. The integral equation was:

$$N(r) = c \int dr'\, N(r')\, F(r') K(|r - r'|)$$

where r is in one or more phase-space dimensions, F(r) is any general function that is generally non-zero, K is a kernel, and c is an eigenvalue. The integration is carried out where F(r) ≠ 0 and solved for N(r) using the smallest eigenvalue, c. F(r) can be of varying complexity – ranging from a constant value to as complex as a differential equation (e.g., differential cross sections). The kernel, K, often took common analytic forms such as the Milne or Gauss kernels describing radiation flow. N(r) was usually the particle flux. Analytical mathematical solutions to this general equation were being generated during that time. The neutron transport equation can be cast into this form. Simplifying assumptions like treating neutrons as mono-energetic, isotropically scattering in simple geometries, using the same mean free path in all materials make this equation analytically solvable.

In 1942 Frankel and Nelson created an analytic solution to this integral form using what they called the end-point method which they later documented in Los

Alamos reports.[12,13,14] They documented their analytic solutions of this integral equation and its applications to determine critical dimensions, multiplication rates, with or without tampers, and probabilities of detonations.[15,16] They also show how the integral equation reduces to the diffusion equation.[17] Frankel and Klaus Fuchs compared the end-point method to the British variation method, and found that they were generally in agreement.[14] Analytic solutions to the neutron transport equation with all of the simplified assumptions continue to be of use for verification of complicated computer algorithms.[18]

*I.C.4. Diffusion Theory Critical Mass Estimates*

The purpose of Project Y (The Los Alamos Project) was to build a practical military weapon that was light enough to carry on an airplane. The preliminary calculations of the minimum quantity of materials and geometry of such a weapon was done using an ideal arrangement but quickly advanced to model more realistic geometries, materials, and combining different areas of physics. The diffusion theory model of neutron transport played an important role in guiding the early insights to the geometry and mass needed to achieve criticality.

Neutron diffusion theory was used as an approximation to the neutron transport equation during the Manhattan Project. It can be derived from the neutron transport equation by integrating over angle and using Fick's Law. Simple diffusion theory treats neutrons in an infinite medium, at a single energy, and assumes steady-state conditions. The diffusion equation for neutron transport is:

$$\frac{1}{v}\frac{\partial \phi}{\partial t} = \nabla(D\nabla\phi) - \Sigma_a \phi + v\Sigma_f \phi + Q$$

or in steady-state:

$$0 = \nabla^2 \phi - \Sigma_a \phi + v\Sigma_f \phi + Q$$

This equation is the basis of the development of neutronics calculations supporting early Los Alamos efforts. The first term describes the amount of neutron movement in the material via diffusion coefficient, D. It includes a Laplacian operator that describes geometry and determines the amount of neutron leakage via surface curvature. The second term is the loss due to absorption. The last two terms are source terms from fission and external sources.

When Serber gave the first lectures at Los Alamos in April 1943 he showed that using simple diffusion theory, with better boundary conditions extending the distance where the flux went to zero, gave a critical radius of a bare sphere as:[3]

$$r_c = \pi^2 D\tau/(v-1)$$

where $\tau$ is the mean-time between fissions, $v$ is the neutron number from fission, and $D$ is the diffusion coefficient. The simple and early estimate for the critical radius of U235 was 13.5 cm giving a critical mass of 200 kg. This early estimate of critical mass is significantly overestimated, a more exact diffusion theory gave a 2/3 reduction. For U235, this gave $r_c$ = 9 cm with a corresponding critical mass of 60 kg which is much closer to the true value.

The effects of a reflective material surrounding the core would serve to reflect a fraction of neutrons that would otherwise escape and therefore reduce the quantity of fissionable material required for a critical mass. Diffusion theory was applied again and showed that the critical mass was 1/8th as much as for a bare system. However, the approximations in diffusion theory over-estimated the conservation of material masses, when more accurate diffusion theory emerged it indicated a reduction in the reflected critical mass of only 1/4th instead of 1/8th.[3]

*I.C.5. Deterministic Solutions*

In deterministic methods, the neutron transport equation (or an approximation of it, such as diffusion theory) is solved as a differential equation where the spatial, angular, energy, and time variables are discretized. Spatial variables are typically discretized by meshing the geometry into many small regions. Angular variables can be discretized by discrete ordinates and weighting quadrature sets (used in the Sn methods) or by functional expansion methods such as spherical harmonics (used in Pn methods). Energy variables are typically discretized by the multi-group method, where each energy group uses flux-weighted constant cross sections with the scattering kinematics calculated as energy group-to-group transfer probabilities. The time variable is divided into discrete time steps, where the time derivatives are replaced with difference equations. The numerical solutions are iterative since the discretized variables are solved in sequence until the convergence criteria are satisfied.

As early as the summer of 1943 it was realized that assuming a single neutron velocity was not adequate. A procedure was developed using two or three energy groups for a single material, but it was cumbersome.[19,20] In October 1943 Christy used a three-group approach in predicting the critical mass of the water boiler.[21] Feynman's group found an easier approach in July 1944 by assuming scattered neutrons did not change their velocity. This allowed each energy group to be treated individually by diffusion theory as in multi-group method.[19,20,22]

The spherical harmonic method, introduced by Carson Mark and Robert Marshak, and developed by Roy Glauber in 1944, solved the tamped critical mass problem for spherical symmetry and one-neutron velocity.[22,23] This was quickly extended to a two-energy group method. After the war Bengt Carlson developed a matrix version of the spherical harmonics method.[22]

In March 1945 Serber introduced an approximation for calculating the critical mass of a reflected sphere.[24,25] The core and reflector could each be described by a reflection coefficient. The critical mass was determined by matching the two reflection coefficients. The neutron distribution in the core was approximated by a sinusoidal form, whereas the reflector neutron distribution was taken as an exponential. The critical radius could be found from a relatively simple formula. Alan Herries Wilson (Figure 1) had derived the same formula for spherical reactors in January 1944,[26] pointed out to Serber by Kenneth Case.[24] Hence, the Serber-Wilson method. Case was immediately able to apply the Serber-Wilson to several complicated configurations that had been difficult with the spherical-harmonic method.[27] The Serber-Wilson method was generalized from one neutron energy to multi-group energy.[28] The Serber-Wilson method was the main neutron diffusion method at Los Alamos until it was replaced by Carlson's Sn transport method in the late 1950s.

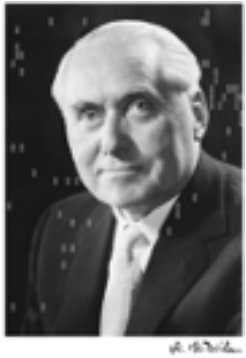

Figure 1. Sir Alan H. Wilson.[29]

In 1953 Carlson first proposed the Sn method, which discretized the angular variables, for solving the transport equation.[31] This was an alternative to the older spherical-harmonic method. The Sn technique is the deterministic method of choice for solving most neutron transport problems today. The technique continues to benefit from past solution techniques to the neutron transport equation. Diffusion Synthetic Acceleration is one example of a technique that was first developed at Los Alamos where diffusion solutions are used to accelerate the convergence of the discrete-ordinates Sn numerical solutions.[32]

*I.C.6. Difficulties Solving Neutron Transport Calculations*

The early estimates of critical dimensions of simple geometries had wide ranging calculated values.[11] The difficulties were from a combination of large uncertainties in nuclear data and limitations in the theoretical description of neutron transport. These limitations were compounded when considering the conditions where geometries are not analytic and materials are undergoing shock from neutron fission heating similar to a pulsed critical assembly.

All the diffusion methods described above were solved numerically by the T-5 desktop calculator group. Solving these complicated physics problems using the T-5 desktop mechanical computers was an additional complication that was tackled by a unique and nearly exclusively groups of women, many of who were the wives of the Los Alamos scientists, using desk calculators. Each person would hand compute segments of a calculation, passing off the resulting value to the next person for further calculation. See the companion articles for further descriptions of the computing.[33,34] The question remained: how could the neutron diffusion equations be solved both quickly and accurately.

By July, 1944, multi-group neutron diffusion algorithms were shown to adequately describe the neutron spatial and energy distribution to enable accurate calculations of critical masses. Approximate formulas for performance and efficiencies were developed by Bethe and Feynman. The laboratory scientists were convinced that they had accurate predictions for critical systems.

At the same time, the laboratory changed focus to systems where fissionable materials are shocked and compressed creating an uncontrolled fission chain reaction resulting in a near-instantaneous release of fission energy. This sudden fission energy release results in a subsequent explosion. Predicting and modeling this process requires detailed understanding of hydrodynamics, material equations of state at the resulting high temperatures and pressures, and neutronics. Equations modeling these complex physics in realistic geometries and materials had to be done accurately.

## II. POST WORLD WAR II: EXPANDED COMPUTATIONAL CAPABILITY

Laboratory scientists became interested in the complexity of the physics, the mathematical and theoretical development, nuclear cross-section and equation of state development, and the computational capability needed to model complex multi-phsyics problems.[33,34,35]

### *II.A. ENIAC – The First Electronic Computer*

Many of the Los Alamos Manhattan Project scientists had left the laboratory after the end of WWII, going to a number of universities across the country. Many, however, remained consultants to Los Alamos. New computing machines were being developed by a number of universities across the country. A strong collaboration naturally existed between the laboratory and universities through the scientists who had worked at both. Many of the scientists knew the priorities at the laboratory were principally centered on improving the complexity of multi-physics calculations.

John von Neumann (Figure 2) was a mathematician and physicist who consulted at Los Alamos on solving key issues. In addition to his major theoretical, mathematical, and physics contributions to improving national defense, von Neumann also encouraged electro-mechanical computing to help solve these complex problems during the wartime effort. Von Neumann's post-war interests continued and played a critical role in introducing electronic computing to Los Alamos. He had frequent communications with his fellow Los Alamos scientists including Stanislaw Ulam and Nicholas Metropolis describing progress in large-scale computing and developing architectures.

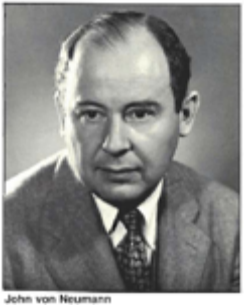

Figure 2. John von Neumann.

As is well known, Herman Goldstine, a mathematician and computer scientist, introduced von Neumann to the ENAIC, Electronic Numeric Integrator and Computer, in the summer of 1944.[33,36,37]

The ENIAC is generally acknowledged as the first electronic, general-purpose digital computer. It consisted of 18,000 vacuum tubes, 70,000 resistors, 10,000 capacitors,1,500 relays, 6,000 manual switches and 5 million soldered joints. It was eight feet high, it covered 1800 square feet of floor space, weighed 30 tons and consumed 150 kilowatts per hour of electrical power. The total cost of the ENIAC was about $7,200,000 in today's dollars.

Although the ENIAC was not built as a stored-program computer, it was flexible enough that it eventually evolved into one.[36,38] The ENIAC was best described as a collection of electronic adding machines and other arithmetic units, which were originally controlled by a web of large electrical cables.[39] The ENIAC was far more difficult to "program" than modern computers. There were no operating systems, programming languages, or compilers available. It was programmed by a combination of plugboard wiring and three portable function tables. Each function table has 1200 ten-way switches, used for entering tables of numbers. There were about 40 plugboards, each several feet in size. A number of wires had to be plugged in for each single instruction of a problem. A typical problem required thousands of connections that took several days to do and many more days to check out.

In 1945, the Army circulated a call for "computers" for a new job with a secret machine (the ENIAC). There was a severe shortage of male engineers during WWII, so six women with science backgrounds were selected to program the ENIAC to solve specific problems. There were no manuals on how to program the ENIAC and no classes were available. They had to teach themselves how the ENIAC worked from the logical diagrams that were available. They figured out how to program it by breaking an algorithm into small tasks and then combining these tasks in the correct order. They invented program flow charts, created programming sheets, wrote programs, and then programmed the ENIAC. The Army never introduced the ENIAC women. It was not until recently that these women received the recognition that they deserved.[37] Without their pioneering programming work, no ENIAC ballistics calculations would have been possible.[36]

### *II.B. ENIAC and Los Alamos*

Construction of ENIAC was not completed until after September 1945 when the war ended. The machine remained significantly untested but the Ballistics Research Laboratory still wanted its firing tables. By December 1945, ENIAC was ready for applications to solve complex problems. A complicated nuclear weapons application from Los Alamos were selected for the first use of ENIAC. The "Los Alamos" problem was to determine the amount of tritium needed to ignite a self-sustaining fusion reaction for the feasibility of the "Super". Experimentation was not possible as there was no tritium stockpile and many years would be needed to produce the inventory of tritium needed for exploration of the conditions for ignition. Nicholas Metropolis and Stanley Frankel first visited ENIAC in the summer of 1945 to understand how it worked. The first calculations, while simplified to lower the security classification, used nearly all of ENIAC's capabilities and were performed in December 1945. After these calculations, the security classification of ENAIC was lowered and ENIAC was announced to the world. On February 15, 1946, the Army revealed the existence of ENIAC to the public (Figure 3).

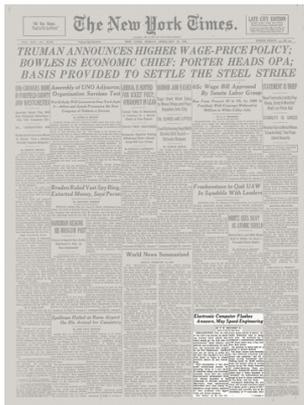

Figure 3. New York Times 15 February 1946 - ENIAC Revealed (need permission to reproduce).

The invention of ENIAC together with national security missions drove a post-war revolution in neutron transport methods development.

## III. THE INVENTION OF THE MODERN MONTE CARLO METHOD

### *III.A. Early History of Statistical Sampling*

The term "Monte Carlo" was used in internal Los Alamos reports as early as August 1947,[40] but appeared publicly for the first time in 1949 with the publication of "The Monte Carlo Method" in the Journal of the American Statistical Association and was written by Nicholas Metropolis (Figure 4) and Stanislaw Ulam.[41] Metropolis humorously relates how the Monte Carlo name came about, as well as the core of the method, which stands today:

*"It was at that time that I suggested an obvious name for the statistical method- a suggestion not unrelated to the fact that Stan had an uncle who would borrow money from relatives because he 'just had to go to Monte Carlo,' The name seems to have endured."*[38]

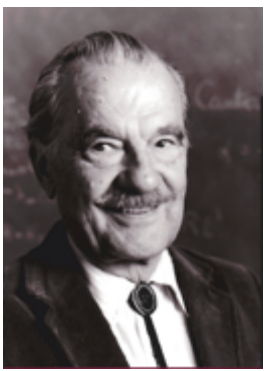

Figure 4. Nicholas Metropolis.

The method can be called a method of statistical trials and has its roots in probability theory that started in the 16th century (Book of Games of Chance, Cardano 1501-1576), and continued with contributors such as Bernoulli (1654–1705), Pascal, de Moivre, Euler, Laplace, Gauss, and Poisson roughly covering the period of 1500–1900 AD. A notable application of these statistical processes was an estimate of the value of π by George Louis Leclerc, also known as Le Comte de Buffon in 1777. The value of π was statistically estimated by repeatedly dropping a needle of known length L on a grid of parallel lines with known spacing, D, with L<D. The value of π was estimated by determining the probability of the needle intersecting the parallel lines using ratios of basic geometric relationships of the needle orientation compared to the line spacing. Buffon obtained the probability of intersecting a line to be: p = 2L / π D. He attempted to calculate the probability, p, by dropping a needle a number of times on a grid and observing the fraction of intersections from the total number of trials. Pierre Simon Laplace (1812) added a novel twist to this idea and examined the special case where L = D. The intersection probability becomes: p = 1 / π. This implies that the ratio of the number of intersections to the number of trials was 1/ π. The application of statistical sampling methods continued by many others, notably Lord Raleigh, who used one-dimensional random walk to provide a solution to a parabolic differential equation. Kolomogorov showed in 1931 the relationship between Markov chain stochastic processes and the solution to some integro-differential equations.

### III.B. Modern Monte Carlo Invented in Los Alamos

The modern Monte Carlo method was invented by Stanislaw Ulam (Figure 5) in mid 1946. The Monte Carlo method is a numerical method for solving stochastic problems by using random sampling. It can be described as a physics experiment carried out numerically on a computer rather than in a laboratory. It involves calculating the average or probability of behaviors of a system by observing the outcomes of a large number of trials that describe the physical events being modeled. Each trial is simulated on a computer according to the evaluation of a series of random numbers. Ulam was a wartime scientist in Los Alamos who had an extensive mathematical background. He relaxed playing solitaire and was intrigued by the theory of branching processes similar to neutron multiplication in fission systems.

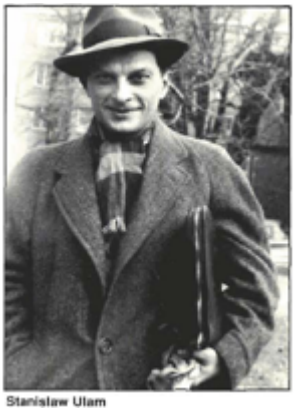

Figure 5. Stanislaw Ulam.

*"The first thoughts and attempts I made to practice [the Monte Carlo method] were suggested by a question which occurred to me in 1946 as I was convalescing from*

*an illness and playing solitaires. The question was what are the chances that a Canfield solitaire laid out with 52 cards will come out successfully? After spending a lot of time trying to estimate them by pure combinatorial calculations, I wondered whether a more practical method than 'abstract thinking' might not be to lay it out say one hundred times and simply observe and count the number of successful plays. This was already possible to envisage with the beginning of the new era of fast computers, and I immediately thought of problems of neutron diffusion and other questions of mathematical physics, and more generally how to change processes described by certain differential equations into an equivalent form interpretable as a succession of random operations. Later. . . [ in 1946, I ] described the idea to John von Neumann and we began to plan actual calculation."*[42]

> THE INSTITUTE FOR ADVANCED STUDY
> PRINCETON, NEW JERSEY
> School of Mathematics
>
> March 11, 1947
>
> Mr. R. Richtmyer
> Post Office Box 1663
> Santa Fe, New Mexico
>
> DEAR BOB,
>   This is the letter I promised you in the course of our telephone conversation on Friday, March 7th.
>   I have been thinking a good deal about the possibility of using statistical methods to solve neutron diffusion and multiplication problems, in accordance with the principle suggested by Stan Ulam. The more I think about this, the more I become convinced that the idea has great merit. My present conclusions and expectations can be summarized as follows:

Figure 6. Portion of Von Neumann's letter to Richtmyer (1947).

Von Neumann saw the significance of Ulam's suggestion and sent a handwritten letter (Figure 6) to Robert Richtmyer, T-Division Leader at Los Alamos, describing the method.[43] The letter contained an 81-step pseudo code for using statistical sampling to model neutron transport. Von Neumann's assumptions were time-dependent, step-wise continuous-energy neutron transport in spherical but radially-varying geometries with one fissionable material. Scattered and fission neutrons were produced isotropically in angle. Fission production multiplicities were 2, 3, or 4 neutrons. Each neutron carried information such as the material zone number, radial position, direction, velocity, and time and the "necessary random values" needed to determine the next step in the particle history after collision. These "tallies" were path length, type of collision, post-collision exit velocity and direction, etc. A new neutron particle (history) was started when the neutron was scattered or moved to a new material. Several neutron histories were started if the collision resulted in a fission reaction. Von Neumann suggested using 100 starting neutrons, each to be run for 100 collisions, and thought that this may be enough. He estimated the time to be 5 hours of computational time on ENIAC. The neutron multiplication rate was the principal quantity of interest for these calculations.

> April 2, 1947
>
> Professor John vonNeumann,
> The Institute for Advanced Study,
> School of Mathematics
> Princeton, New Jersey
>
> Dear Johnny:
>
> As Stan told you, your letter has aroused a great deal of interest here. We have had a number of discussions of your method and Bengt Carlson has even set to work to test it out by hand calculation in a simple case.

Figure 7: Portion of Richtmyer's response to von Neumann (1947).

Richtmyer responded (Figure 7) conveying that he was very interested in the idea but made some suggested changes to von Neumann's assumptions. Richtmyer suggested allowing for multiple fissionable materials, no fission energy-spectrum dependence, single neutron multiplicity, and to run for computer time and not collisions. The coding for ENIAC was finalized in December 1947 and contained significant advances in neutron transport including step-wise continuous-energy neutron cross section data, fission spectra, and nuclear cross sections tabulated at interval mid-points with histogram energy-dependence. The first-ever pseudo-random number generator was also created. The first Monte Carlo calculations were carried out on ENIAC in April/May 1948.

### III.C. Enrico Fermi and the FERMIAC

In 1947, when the ENIAC was out of service while being moved to the Ballistics Research Laboratory, Enrico Fermi invented a 30-cm long, hand-operated, analog, mechanical computer that was used to trace the histories of neutrons moving through materials using the Monte Carlo method. The machine, called the FERMIAC, was constructed by L.D.P. King. King worked at Omega Site in Los Alamos with the Water Boiler Reactor. Enrico Fermi was a part of the Los Alamos staff during the war and returned to the University of Chicago postwar but remained a consultant to Los Alamos and frequently visited. In 1947, he found his fellow scientists frustrated with having a novel method for solving neutron diffusion (Monte Carlo) and not having an appropriate computer to use it. ENIAC was being moved from the University of Pennsylvania to Aberdeen Proving Grounds. During one visit, Fermi discussed creating a mechanical device to follow neutron tracks through materials with King at a family picnic. He explained the Monte Carlo process and how the device would work to King. King spent some time with Fermi in the Omega site workshop to design the mechanical Monte Carlo computer. The result was a trolley with a 10-step brass roller which would roll out the neutron range in the given material. The FERMIAC required a scaled drawing of the nuclear device, physics data tables, and random number lists. After the initial collection of source neutrons were decided on, a mark on the scaled drawing was made to note the precise point of collision to determine the next flight path and angle. Elapsed time was based on the neutron velocity and was grouped as "fast" or "slow". Distance traveled between collisions was mechanically measured based on the neutron velocity and material properties.

Bengt Carlson's T-Division Group at Los Alamos used the FERMIAC from 1947 to 1949 to model various nuclear systems with at least 100 source neutrons. An example of one of its uses was to determine the change in neutron population with time.

An increasing neutron population would represent a supercritical system and a decreasing population would represent a subcritical system. A steady neutron population indicated a critical system. Bengt Carlson said that "although the trolley itself had become obsolete by 1949, the lessons learned from it were invaluable."[44]

Recently, the Enrico Fermi Center in Italy became interested in the FERMIAC. Using the original Los Alamos FERMIAC drawings, a replica was built at the Italian research agency National Institute for Nuclear Physics in 2015. They recreated the procedure for using the FERMIAC. A simulation using the replica was made for an air-cooled, uranium graphite reactor. Figure 9 shows the results of the tracks from 100 source neutrons. The three regions are fuel, air, and graphite. Those neutrons that sample a fission reaction spawn multiple tracks.[45]

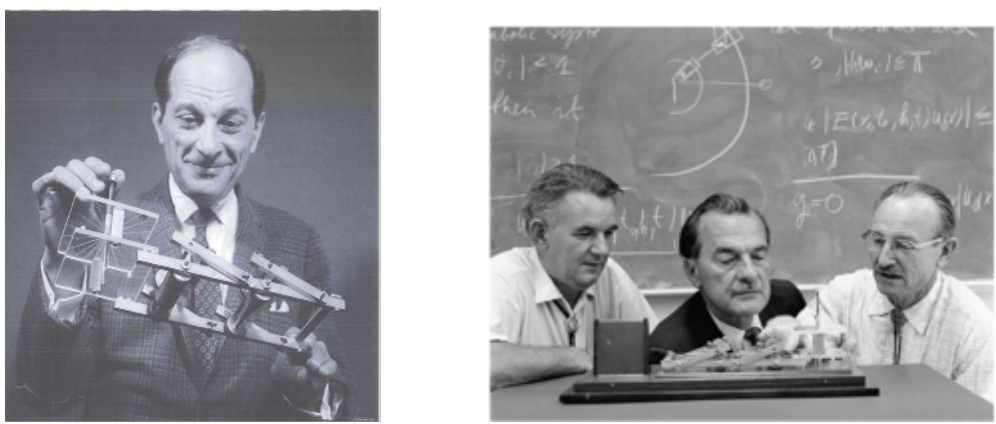

Figure 8. Stan Ulam, Bengt Carlson, Nicholas Metropolis, and L.D.P King with FERMIAC (1966).[46]

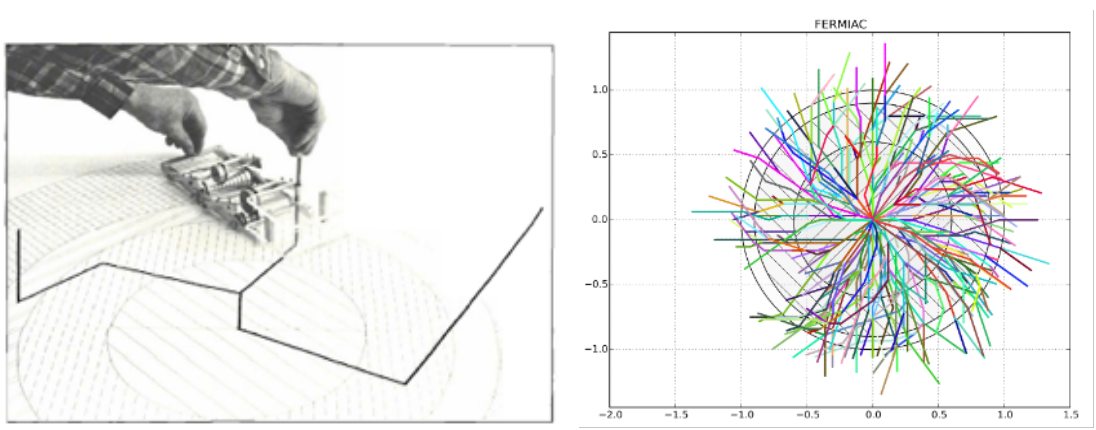

Figure 9. FERMIAC tracking particles.[47]

### III.D. Programming the First Monte Carlo Calculations on the ENIAC

The first ENIAC Monte Carlo calculations were chain reaction simulations for Los Alamos and occurred in three separate groups of problems: April–May 1948, October–November 1948, and May–June 1949. There were Monte Carlo calculations run for Argonne for nuclear reactor calculations in December 1948,[36] that used the same basic Monte Carlo algorithm as in the two previous calculations. These ENIAC calculations were all performed at the Army's Aberdeen Proving Ground in Maryland.

The Los Alamos model for performing calculations on the ENIAC was a two-step process of planning and coding.[36] The planning was the theoretical development of the algorithm to solve the problem being considered. The coding was teaching or programming the ENIAC how execute the algorithm.

John Von Neumann did the planning to create the algorithm to solve the neutron transport problem. Flow diagrams of the algorithm were considered crucial to providing a rigorous approach for the translation of the mathematical expressions into machine language programs. Two complete Monte Carlo flow diagrams from 1947 have been preserved.[36]

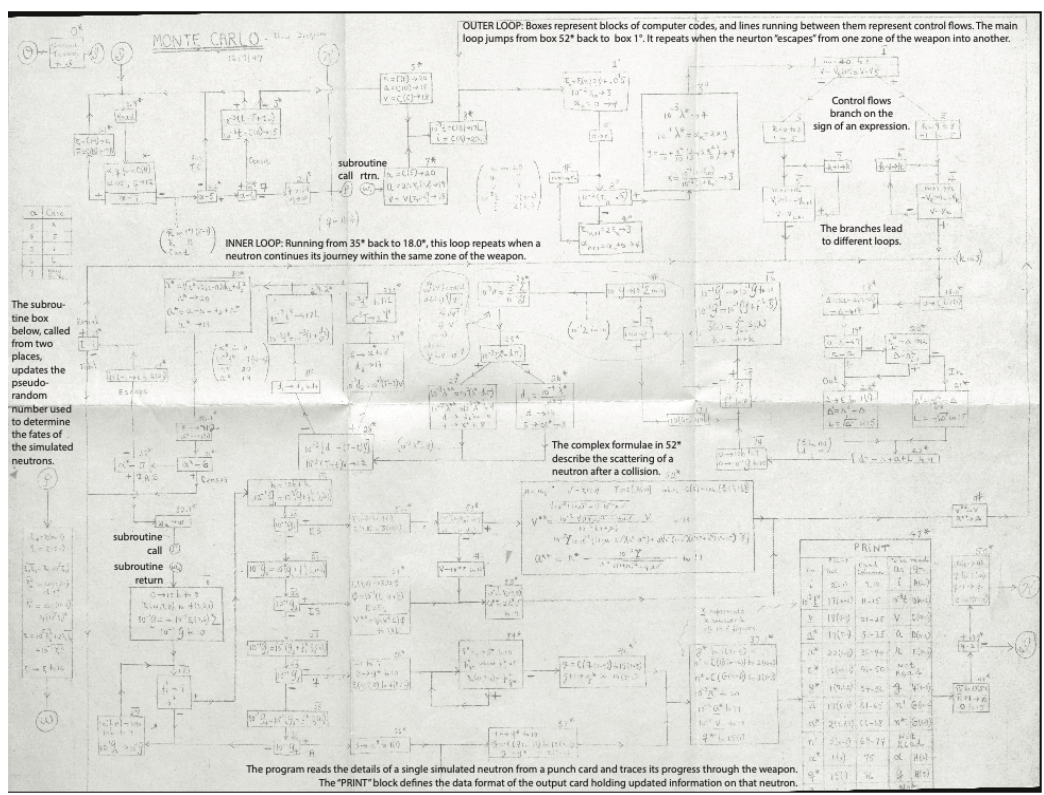

Figure 10. Programming flow chart for Monte Carlo calculation on ENIAC (need permission to reproduce).[36]

Before coding could begin, they and the BRL staff converted ENIAC into a stored program computer, programming via patch cables would no longer be required.[36]

The coding of the Monte Carlo run was performed by both Klara Von Neuman and Nicholas Metropolis.[36] Much of the responsibility for diagramming and coding the Monte Carlo algorithm belonged to Klara Von Neumann. She described programming as being "just like a very amusing and rather intricate jigsaw puzzle, with the added incentive that it was doing something useful." For 32 days straight in April 1948, they installed the new control system, checked the code, and ran ENIAC day and night. Klara was "very run-down after the siege in Aberdeen and had lost 15 pounds." Programming and running the ENIAC was not an easy task.[36]

Klara von Neumann documented the techniques used to program the Monte Carlo algorithm in a manuscript "III: Actual Technique—The Use of the ENIAC."[36] This document began with a discussion of the conversion of the ENIAC to support the new code paradigm, documented the data format used on the punched cards, and outlined in reasonable detail the overall structure of the computation and the operations performed at each stage. An expanded and updated version of this report was written by Klara von Neumann and edited in collaboration with Nick Metropolis and John von

Neumann during September 1949. It contains a detailed description of the computations, highlighting the changes in the flow diagram, program code, and manual procedures between the two versions.[36]

The first Monte Carlo problems included "complex geometries and realistic neutron-velocity spectra that were handled easily."[38] The algorithm used for the ENIAC is similar in many respects to present-day Monte Carlo neutronics codes: read the neutron's characteristics, find its velocity, calculate the distance to boundary, calculate the neutron cross section, determine if the neutron reached time census, determine if the neutron escaped, refresh the random number, and determine the collision type.[36] A card would be punched at the end of each neutron history and the main loop restarted. The neutron history punched cards were then subjected to a variety of statistical analyses.[38]

The 840-instruction Monte Carlo program on the ENIAC included a number of firsts for computer programming. "The program included a number of key features of the modern code paradigm. It was composed of instructions written using a small number of operation codes, some of which were followed by additional arguments. Conditional and unconditional jumps were used to transfer control between different parts of the program. Instructions and data shared a single address space, and loops were combined with index variables to iterate through values stored in tables. A subroutine was called from more than one point in the program, with the return address stored and used to jump back to the appropriate place on completion. This was the first program ever executed to incorporate these other key features of the modern code paradigm."[36]

The ENIAC Monte Carlo calculations were extremely successful.[36,38] The unique combination of national security needs, brilliant scientists, the first programmable computer, and very clever programmers of ENIAC made these results possible. Today's Monte Carlo codes build on these first ENIAC results.

### *III.E. Generation and Use of Random Numbers on the ENIAC*

*"This meant generating individual neutron histories as dictated by random mechanisms-a difficult problem in itself,"* Metropolis explained. *"Thus, neutrons could have different velocities and could pass through a variety of heterogeneous materials of any shape or density. Each neutron could be absorbed, cause fission or change its velocity or direction according to predetermined computed or measured probability. In essence, the 'fate' of a large number of neutrons could be followed in detail from birth to capture or escape from a given system."*[44]

A "Monte Carlo" calculation has been generally defined as one that makes explicit use of random numbers. Nature is full of completely random processes such as neutron transport, the number and energy of neutrons emitted from nuclear fission, and radioactive decay. For example, if a neutron is born in an infinite material with a specified energy, location, and direction, the neutron will undergo various interactions at different locations in the material. If a second neutron is born in the material with the same energy, location, and direction, that neutron will have a different set of interactions at different locations in the material. Each neutron random walk will be different. Any attempt to model these neutron random walks must include a capability to model this random behavior.

### *III.E.1. Random Numbers for Sampling Random Processes*

The availability of random numbers uniformly distributed in the interval (0,1) is absolutely essential to obtain correct results when modeling any random process. These random numbers will be used to sample a probability density function to select the outcome of a process or event. In the case of the neutrons above in an infinite material, random numbers would be used to sample a distance to collision, the isotope that was collided with, the type of collision that occurred (e.g., scatter, fission, or absorption), and the energy and direction of the neutron(s) after collision if it is not absorbed.

Before the advent of electronic digital computers, random numbers were produced in the form of tables that were used as needed for hand calculations. Entire books were published containing only random numbers. As easily imagined, this method of looking up random numbers by hand was very tedious and time consuming.

When electronic digital computers became available, a different method for "looking up" random numbers could be used. Deterministic algorithms were devised that generated a sequence of pseudo-random numbers. The term "pseudo" indicates that the deterministic algorithm can repeat the sequence and therefore the random numbers are not truly random: they imitate randomness. In this paper, the term "random number" means pseudo-random number. Tests for randomness - no observed patterns or regularities - were developed to assess how random these deterministic sequences appeared to be.

*III.E.2. Random Number Generation and Use by the ENIAC*

When John Von Neumann was developing the neutron diffusion algorithm (involving random processes) for the ENIAC, he considered the best way to have random numbers available for use in the algorithm. The amount of memory available on the ENAIC was small and could not be used to store a table of random numbers. He realized that having the ENAIC itself create the random numbers during the neutron diffusion calculation was going to be much faster than reading them from punched cards. He developed a Random Number Generator (RNG) for the ENIAC that appeared to generate acceptable random number sequences and was easy to implement. A famous 1949 quote from John Von Neumann is "Anyone who attempts to generate random numbers by deterministic means is, of course, living in a state of sin." His quote is to warn users that these (pseudo-)random numbers are not truly random. However, he definitely favored the use of these random numbers for modeling random processes as he himself did on the ENIAC.

John Von Neumann invented the middle-square method RNG in 1946.[48] An n-digit number (n must even) is squared, resulting in a 2n-digit number (lead zeros are added if necessary, to make the squared number 2n digits). The middle n digits are used as the random number. The next "refreshed" random number takes the last random number and squares it. This process is repeated whenever a new random number is needed. Values of n used on the ENIAC were 8 and 10 during the first Monte Carlo calculations. There were statistical tests for randomness that were performed by hand on the first 3000 numbers generated by this scheme. These tests seemed to indicate that the numbers appeared to be random.[36] It turns out that the middle-square RNG is not a good one because its period is short and it is possible that this RNG can get stuck on zero. It was adequate, however, for the ENIAC runs that could only calculate a small number of neutron histories and therefore, only required a small number of random numbers.

The code for this RNG in the ENIAC program flow diagram was at first replicated wherever a random number was required. In late 1947, a different programming paradigm was used where the RNG was its own entity and was entered from different locations in the Monte Carlo neutron diffusion program. This technique of reusing code was one of the first calls to a subroutine in computer programming. The result of this work was a series of successful Monte Carlo neutron diffusion calculations.[36]

## IV. MONTE CARLO METHODS IMPROVEMENTS ARE REQUIRED (1950–PRESENT)

As the interest in the Monte Carlo method grew after the successful ENIAC experiences, many organizations began to study and develop new or modified Monte Carlo methods for their own applications. Today, Monte Carlo methods are used to solve problems in many disciplines including the physical sciences, biology, mathematics, engineering, weather forecasting, artificial intelligence, finance, and cinema. This discussion will focus on the Los Alamos Monte Carlo developments for neutronics calculations below 20 MeV since the ENIAC calculations.

Monte Carlo solutions are usually associated with the integral form of the neutron transport equation, which can be derived from the integro-differential form.[49] It is called the linearized Boltzmann transport equation and is written as:

$$\frac{1}{v}\frac{\partial \psi}{\partial t} = Q + \int d\Omega' \int dE' \Sigma_s \psi + \frac{\chi}{4\pi} \int d\Omega' \int dE' \nu \Sigma_f \psi - (\Omega \cdot \nabla + \Sigma_t)\psi$$

This is a balance equation (gains minus losses) where the time-rate of change in particle flux is equal to the external source, scattering, and multiplication (gains) minus the leakage and collisions (losses). This equation can be solved directly by Monte Carlo. Particle history random walks are sampled one at a time from probability density functions involving measured physical data. Each particle history is an independent sample and is only dependent on the last collision event. The Monte Carlo results (tallies) are integrals over all the random walks to produce an average answer for the desired quantity. Monte Carlo results are confidence intervals formed by applying the independent- and identically-distributed Central Limit Theorem.[50]

Monte Carlo methods have the distinct advantage of being able to sample all the phase-space variables as continuous functions and not discretized functions. The disadvantage is that the estimated statistical error in a tally result decreases as 1/sqrt(N), where N is the number of histories run. To reduce the error by two, N must increase by four. Variance reduction methods usually must be used to be able to run efficient Monte Carlo simulations.

The Monte Carlo algorithm was the method of choice for solving complicated neutron diffusion problems in the late 1940s because it was a better model of neutron transport and was fairly easily adapted to the ENIAC. The Monte Carlo principles used on the ENIAC are the same ones applied today.

Modern Monte Carlo developments are discussed below in five areas: theory, RNGs, computational physics, data libraries, and code development. A comprehensive list of Los Alamos Monte Carlo references covering the past eight decades is available online.[50]

### *IV.A. Los Alamos Developments in Monte Carlo Transport Theory*

#### *IV.A.1. General Monte Carlo Theory*

The Atomic Energy Act of 1946 created the Atomic Energy Commission to succeed the Manhattan Project. In 1953, the United States embarked upon the "Atoms for Peace" program with the intent of developing nuclear energy for peaceful applications such as nuclear power generation. Meanwhile, computers were advancing rapidly. These factors led to greater interest in the Monte Carlo method. In 1954 the first comprehensive review of the Monte Carlo method was published by Herman Kahn,[51] and the first book was published by Cashwell and Everett in 1959.[52]

The first Los Alamos comprehensive document discussing the theoretical and practical aspects of Monte Carlo development for neutron (and photon) transport was written in 1957 in a Los Alamos internal report and later published.[52] This report presents in detail all the aspects of a Monte Carlo particle random walk from birth to death. The concept of particle weight is discussed, as well as several types of variance reduction methods such as cell importances, forced collisions, and weight, energy, and time cutoffs. The central limit theorem and the statistical aspects of Monte Carlo confidence interval results are also reviewed. The principles presented here were all incorporated into the first Los Alamos general Monte Carlo code, MCS,[53] and still exist today in the present Monte Carlo code Monte Carlo N-Particle® (MCNP®).[54] The appendix[52] lists 20 problems that were solved on the MANIAC I, the Los Alamos computer that was used after the ENIAC, with the Monte Carlo method.

The next major Los Alamos Monte Carlo review publication was written in 1975 and described many advances in both computational physics and variance reduction techniques.[55] Both fixed source and nuclear criticality $k_{eff}$ eigenvalue problems are discussed. Some of the topics that are included are sampling by rejection, importance sampling, splitting and Russian roulette, the exponential transform, and various tally estimators. The Monte Carlo topics discussed in this document are still in use today in all Monte Carlo radiation transport codes including the MCNP® code.

Since 1975, there have been many major Los Alamos developments in Monte Carlo theory, a few of which are mentioned here. The first is the concept of weight windows for energy-, time-, and space-dependent particle control.[56] Several methods were developed to set the weight windows for a problem including the MCNP® weight window generator and adjoint deterministic solutions. A second concept is the DXTRAN sphere that allows Monte Carlo particles to move to small regions of a problem that are deemed important.[57] Upgrades were made to the random number generator to allow a longer period and a skip-ahead capability that is required for parallel Monte Carlo calculations.[58] A new method to assess the convergence of Monte Carlo $k_{eff}$ eigenvalue calculations was developed, implemented, and tested in the MCNP® code.[59] New statistical procedures were developed to assess the proper convergence of Monte Carlo results.[60] Research into new Monte Carlo methods is an ongoing activity. Stay tuned.

*IV.A.2 Non-Boltzmann Monte Carlo Tallies*

Another new method was created to be able to use variance reduction for non-Boltzmann tallies such as pulse-height tallies.[61] The Monte Carlo solution to the Boltzmann Neutron Transport Equation calculates average contributions to a tally such as flux or current over many particle random walks. These tallies are averages of portions of a random walk where each portion is treated as a separate, independent contribution to a tally. For example, if a particle enters a cell, collides (losing energy), and exits the cell, there would be two contributions to a cell tally: one at each energy.

The Monte Carlo method offers the possibility of calculating other types of tallies that include correlations of tracks and progeny tracks within one source history. These correlated tallies are called non-Boltzmann tallies because the Boltzmann transport equation contains no information about these correlations. Examples of non-Boltzmann tallies would be coincidence counting and pulse height tallies. A pulse height tally is defined as the sum of the energy deposited in a cell from all the contributions of one source particle. The final tally would not be known until the source history and all its progeny have been completed. In the one collision example above, there would now be one contribution for a pulse height tally in the cell instead of two.

The detail that exists in all the tracks of a Monte Carlo random walk allows keeping a record of all of these correlations. As expected, it is much more difficult to sample correlated tallies than uncorrelated tallies, often making efficient non-Boltzmann calculations extremely difficult. New variance reduction methods that apply to a collection of tracks were developed at Los Alamos for these non-Boltzmann class of tallies including deconvolution, corrected single particle, and the supertrack.[61] The deconvolution method was selected to implement into the MCNP® code.[62,63] This algorithm has been fully tested,[64,65] so that efficient pulse height tally calculations using variance reduction can be made.

### *IV.B. Random Number Generators*

There are many different deterministic RNGs that have been developed over the decades since Monte Carlo and other statistical sampling methods have become popular. Before being used, these RNGs are tested for repeatability, robustness, computational efficiency, and random behavior of the calculated random numbers to assess their ability to correctly sample random processes.

The period of a RNG is the number of random numbers that can be generated before the random number sequence repeats itself. As computers became faster and calculations became more complex, RNGs with longer periods were required so that no random numbers are reused during the modeling of the random processes.

There are two types of RNGs available today: hardware-based RNGs that are constantly changing based on some environmental variable(s) and are therefore not repeatable, and the deterministic pseudo RNGs that are repeatable. Pseudo RNGs are the method of choice for Monte Carlo calculations because they are repeatable and therefore can be easily tested and calculations more readily compared.

The pseudo RNGs of today are vastly superior to the middle-square RNG method of ENIAC times. They are more random, are extremely robust, have much longer periods, and allow easy access to any location in the sequence. In addition, the new statistical tests for randomness are more sensitive at finding patterns and regularities. A detailed discussion of all the RNGs available today is beyond the scope of this paper. What follows is a brief summary of the RNGs available for use in neutronics calculations by the Los Alamos Monte Carlo N-Particle® transport code MCNP®.[66]

The big improvement in RNGs was the Linear Congruential Generator (LCG) that first appeared in 1951[67] and was later improved in 1958.[68] LCGs are the oldest, best known, and most studied RNGs. Each random number is calculated based on a recurrence relation involving the previous random number. The initial random number is called the seed. LCGs have many desirable characteristics for Monte Carlo calculations.[58] They have been extensively studied and tested. They are repeatable, have long, known periods, are robust, require little computer memory, and are computationally very efficient. It is easy to skip ahead to any location in the sequence so that the i_th source particle always starts with the same sequence for the specified LCG seed. This feature enables correlated sampling comparisons between similar calculations. This skip-ahead capability is also crucial for performing parallel Monte Carlo calculations. For these reasons, the LCG is used in many Monte Carlo radiation transport codes including the MCNP® code.

The MCNP® code uses the LCG with 48-bits-of-precision integer arithmetic as the default RNG. This RNG has had over 50 years of continuous use in MCNP and its predecessors on many different computers and compilers. It has been proven to be repeatable, extremely reliable, robust, random, and has a long period of 2**46 (10**14 random numbers). This RNG algorithm has been made completely portable and has been adapted to many different computing environments.

The default MCNP® LCG was upgraded and extensively retested in 2007.[58] More recently, 63-bit LCGs have become available in MCNP®.[69,70] These 63-bit LCGs have a period of up to 2**63 (10**19 random numbers) that is 130,000 times longer than the 48-bit LCGs. It is interesting to note that the default random number stride, the number of random numbers between source particles, in the 1960's was 4297. The current default stride was changed to 152917 in 1990. The longer stride is required because the MCNP® problems run today are far more complex than in decades past and thus require more random numbers. The MCNP® code allows the user to select which one of 13 RNGs to use, the starting random number seed, the stride between source particles, and the history number of the first source particle (usually for studying details of a specific particle random walk).

Since high quality RNGs are absolutely essential for obtaining good numerical solutions to problems involving random processes, performing RNG testing is a crucial activity. Over the years, the MCNP® 48-bit RNG has been tested many times and in various ways. A detailed and thorough statistical test study of the current MCNP® 48-bit LCG and 12 63-bit LCG RNGs was performed in 2011.[58] A total of nearly thirty theoretical and empirical tests were made for each RNG. The study concludes that the current MCNP® RNG exhibits excellent random behavior, and the next longer period RNG for the MCNP® code should most likely be a L'Ecuyer 63-bit LCG.[70]

*IV.C. Los Alamos Developments in Neutron Computational Physics for MCNP*

The early neutron computational physics algorithms used simple models and data libraries that were not very detailed. Only fixed source problems were initially considered. New algorithms and estimators were developed for nuclear criticality problems to solve for the unknown spatial fission distribution and the $k_{eff}$ eigenvalue. The only thermal neutron treatment available was the free gas model where the neutrons are assumed to be transported in a monatomic gas with a Maxwellian distribution of velocities.[53] Time-dependent material temperatures were allowed using the free gas model.

The computational physics improvements in the MCNP® code involve using improved models and more complete data libraries. These new libraries use different data laws to more correctly represent neutron (n,xn) reactions, neutron fission multiplicities, prompt and delayed fission neutron production, other neutron-producing reactions such as photofission, and cross-section uncertainties. There are also new thermal neutron data for the double-differential S(a,b) treatments that include temperature-dependent molecular and crystalline effects. A new algorithm was developed to treat unresolved neutron resonances.[71] On the-fly Doppler broadening has been implemented to include temperature effects on the neutron cross sections during the calculation.[72] There are many other improvements not only in physics, but also in the code architecture, geometry, tallies, graphics, and user features to aid in providing an understanding of the problem results that are described in Ref. 50.

*IV.D. Los Alamos Monte Carlo Code Development History*

The history of Monte Carlo code development began with the John von Neumann neutron diffusion code for the ENIAC. Every Los Alamos computer has run John von Neumann-types of Monte Carlo calculations. The timeline for the development of various Los Alamos Monte Carlo codes is thoroughly discussed in Ref. 73. A short synopsis of the code history is presented here.

At Los Alamos, Monte Carlo computer codes developed along with computers. The first Monte Carlo code was the simple 19-step computing sheet in John von Neumann's letter to Richtmyer.[43] But as computers became more sophisticated, so did the codes. At first, the codes were written in machine language and each code would solve a specific problem. In the early 1960s, better computers and the standardization of programming languages such as FORTRAN made possible more general codes.

The first Los Alamos general-purpose Monte Carlo particle transport code was MCS,[53] written in 1963. MCS was programmed in the FLOCO coding system for the IBM 7090 computer and was used for neutron shielding calculations for a time-independent geometry. It was capable of treating an arbitrary three-dimensional configuration of first- and second-degree surfaces. Scientists who were not necessarily experts in computers and Monte Carlo mathematical techniques now could take advantage of the Monte Carlo method for solving radiation transport problems. They could run the MCS code to solve modest problems without having to do either the programming or the mathematical analysis themselves.

MCS was followed by MCN[74] in 1972 and was written in FORTRAN IV. MCN could solve neutron transport problems in a user-defined three−dimensional geometry created using 24 surface types. MCN had six standard sources and a user-defined subroutine source. There were six standard tallies including a point detector tally. Only one tally of each type was allowed. The neutron cross sections were accessed from data stored in separate libraries from either Livermore or Aldermaston. The basic MCN input file description is the same one that is used in MCNP today.

In 1977, MCN and other Los Alamos Monte Carlo transport codes were merged together and named MCNP.[75] At first MCNP stood for Monte Carlo Neutron Photon. Today it stands for the Monte Carlo N−Particle® code since high-energy physics particles are now available. The MCNP® code now included the present generalized tally structure, automatic calculation of volumes, and a Monte Carlo eigenvalue algorithm to determine $k_{eff}$ for nuclear criticality calculations.

In 1983 MCNP3 (Version 3) was released. It was entirely rewritten in ANSI standard Fortran 77. MCNP3 was the first MCNP version internationally distributed through the Radiation Safety Information Computational Center at Oak Ridge, Tennessee. Other 1980s versions of MCNP were MCNP3A (1986) and MCNP3B (1988), that included tally plotting graphics, the present generalized source, surface sources, repeated structures/lattice geometries, and multigroup/adjoint transport.

MCNP4 (1990) and was the first multitasking version of the code. MCNP4 added electron transport, the pulse height tally, a thick−target bremsstrahlung source approximation for photons, enabled detectors and DXTRAN with the S($\alpha,\beta$) thermal treatment, and provided greater random number control. MCNP4A (1993) featured enhanced statistical analysis, distributed processor multitasking for running in parallel on a cluster of scientific workstations, new photon libraries, ENDF−6 data capabilities, color X−Windows graphics, dynamic memory allocation, expanded criticality output, periodic boundaries, plotting of particle tracks, and improved tallies in repeated structures. MCNP4B (1997) featured differential operator perturbations, enhanced photon physics, PVM load balance and fault tolerance, cross-section plotting, postscript file plotting, 64−bit workstation upgrades, PC X−windows, lattice universe mapping, enhanced neutron lifetimes, and a coincident−surface lattice capability. MCNP4C (2000) featured an unresolved resonance treatment, macrobodies, superimposed importance mesh, perturbation enhancements, electron physics enhancements, plotter upgrades, cumulative tallies, and parallel computing enhancements.

MCNP5 (2003) was rewritten in ANSI standard Fortran 90 and includes the addition of photonuclear collision physics, superimposed mesh tallies, time and energy splitting, radiography tallies, and plotter upgrades. MCNP5 also included parallel computing enhancements with the addition of support for OpenMP and MPI. MCNP5 Version 1.40 (2005) added lethargy plots, logarithmic data interpolation, neutron multiplicity distributions from fission, stochastic geometry, source entropy, mesh tally plots, new electron energy loss straggling. MCNP5 Version 1.15 (2009) includes photon Doppler broadening, variance reduction with pulse height tallies, annihilation gamma-ray tracking, and large lattice enhancements. MCNP5 Version 1.60 (2010) adds adjoint weighted tallies for point kinetics parameters, mesh tallies for isotopic reaction rates, up to 100 million cells and surfaces, and up to 10 thousand tallies.

MCNP6 Version 1.0 (2013) is a merger of the MCNP® and MCNPX codes.[76] MCNP6.1 can transport 37 different particles to energies greater than 1 TeV.[50] MCNP6.1 also contains unstructured mesh geometry transport,[77] automatic weight-window generation with SN code PARTISN, photon transport to 1 eV, magnetic field tracking in air, new depletion capabilities, parallel MPI improvements, cosmic ray sources, ENDF/B-VII.1 data, 64 bit Windows plotting, beta decay source, new pulse height tally options, full use of continuous S(a,b), and keff perturbations. MCNP6.2 (2018) contains many improvements in physics, sources, tallies, unstructured mesh, and data such as correlated prompt fission neutron and gamma-ray emission models, improved cosmic-ray source, new default cross sections, built-in physics-based neutron and photon response functions, and improved tracking of charged particles on an unstructured mesh. 2017 marked the 40[th] anniversary of the MCNP® code and the 70[th] anniversary of the invention of modern Monte Carlo. The same year, the laboratory was awarded registered trademarking of its flagship code. The Monte Carlo N-Particle ® and MCNP® names were granted registered trademarks by the US Patent and Trademark Office and are the only two registered trademarks of Los Alamos National Laboratory. Several new software packages are available to use with the MCNP® code.[54] The Whisper code is used for sensitivity-uncertainty-based nuclear criticality safety validation.[78,79] MCNPTools is a collection of commonly used utilities that perform functions such as merging files from different calculations.[80] The true power behind MCNPTools is that it allows the user to write their own custom tools and process MCNP® code output. The Intrinsic Source Constructor is a software library and associated data files to construct radioactive source descriptions given a set of material isotopics.[81] In 2018, the MCNP® code development team embarked on a large code modernization process. Large structural changes are currently being made to make the coding computationally efficient for large scale problems and expected computing platforms. These changes are expected to also significantly simplify code maintenance and speed up new capability development time. MCNP6.3 is in development and expected to be released in 2021.

*IV.E. Monte Carlo Neutronic Data Library Development History*

*IV.E.1. Theme for the Past 65 Years—Nuclear Data Is Important and Presents Challenges to Code Developers*

Monte Carlo code development at Los Alamos has been inexorably linked with development and faithful utilization of nuclear data libraries. As noted in 1975 by Carter and Cashwell: *"The ability to perform sophisticated particle-transport calculations is of questionable value unless reliable basic nuclear data are also available. Significant progress has been made in this direction during the past few years, and evaluated nuclear data files are now readily available. This relieves the user of some anxieties about cross sections, but he should always remember that reliable output depends upon good cross-section input."*[55]

The need for accuracy in nuclear data, along with resulting challenges, was noted in the documentation of the first general-purpose Monte Carlo code at Los Alamos, MCS, in 1963: *"In general, the treatment of the nuclear reaction data has been designed to represent accurately the experimental data – frequently at the expense of computing time."*[53]

The first nuclear data was measured in the late 1930s, and the Manhattan Project carried out many of the fundamental nuclear data measurements. Los Alamos continues to this day to make nuclear data measurements. By 1972 compilations of data were starting to become available. The code developers continued to attest to faithful utilization of such data: *"… in MCN we treat the various reactions precisely as they are described in the LLL and the UK compilations …,"*[74] and *"Our aim has been to use the data provided in this code with no introduction of significant alterations or processing of the data by us."*[74]

In a 1975 report on the status of Monte Carlo at Los Alamos,[82] it was noted that the Laboratory's Monte Carlo codes were under continuous development *"in order to meet the increased demands upon then, both as to the difficulty of the problems they are asked to solve and with respect to the task of handling the vast amount of cross-section data now available."* Tension between accuracy and storage requirements were noted: *"The cross sections are read into the codes in considerable detail in an attempt to use the information with no significant approximations or distortions. This puts a considerable burden on the storage capacity of the computer, especially in view of the increasing size of nuclear data compilations such as the Evaluated Nuclear Data Files (ENDF)."*

Finally, in 1980 when the MCNP® code was under early development, a new report on the status of Monte Carlo at Los Alamos asserted that *"MCNP is a continuous-energy Monte Carlo code that makes no gross assumptions regarding data."*[83]

*IV.E.2. Where Did Nuclear Data for Los Alamos Monte Carlo Codes Come From?*

Documentation for the MCS code in 1963 describes cross-section input that is provided to the code, including cross sections, angular distributions, and energy distributions. It is unclear where code users obtained that data, although the document describes it as "experimental data." Cross-section experiments had been carried out at Los Alamos and many other institutions from the time of the Manhattan Project through the early 60s (see Chadwick, this volume. An excellent example of a compendium of early cross-section experimental data can be found in Ref. 84.

By 1972, the MCN code accessed pointwise data in LLL or AWRE format. Interestingly, the data were read from a library maintained on the disk of the MANIAC. Documentation of the UK Atomic Weapons Research Establishment nuclear data library at this time may be found in Ref. 85. The LLL data have been described over the years in a comprehensive series of Lawrence Livermore Reports UCRL-50400, for example, Ref. 86. From Rev. 1 of that document (1981) a bit of the history was presented: "*The LLNL Evaluated Nuclear Data Library has existed since 1958, in a succession of forms and formats. In its earliest form it was a series of internal memos containing tabulations of cross sections and angular distributions for a few isotopes to be used in neutronics calculations. It was soon found that some type of mechanization should be undertaken both for efficiency and for convenience. Thus, in rapid succession, the library went through stages of punched card and punched paper-tape, and then to the first magnetic-tape BCD card image form. Of course, once the library was in one computer-readable form, it could be translated with relative ease to another computer-readable form.*"

By 1975, MCN users had access to ENDF data as well: "*The pointwise cross sections used in MCN at present are from the nuclear data compilations of Howerton's group at LLL, from ENDF, or from the British (AWRE).*" The US national effort for a standard nuclear data compilation (ENDF – Evaluated Nuclear Data File) was led by the Cross Section Evaluation Working Group (CSEWG). A fascinating history of CSEWG and ENDF is found in Ref. 87. The origins of ENDF are described in that document as follows: "*In 1966, the Division of Reactor Development and Technology (DRDT) of the United States Atomic Energy Commission (USAEC) was concerned about the problems involved in the evaluation and processing of nuclear data for reactor calculations. The DRDT's plan for the development of the necessary methods for the processing of the data and for obtaining data for immediate use in reactor calculations involved a long range goal of developing automated methods for processing nuclear data, as well as a short range goal of providing reactor designers with a reference set of data that they could use for their current projects. For the short term goal of obtaining a reference set of nuclear data, the DRDT sponsored the Cross Section Evaluation Working Group (CSEWG), a co-operative evaluation effort aimed at providing reactor designers with a good set of evaluated nuclear data. The proposed long range goal was to be addressed by CSEWG over time.*"

Releases of ENDF libraries are notated by a Roman numeral for major releases, and sometimes by an Arabic number for minor releases (e.g., ENDF/B-VII.1). The original release, ENDF/B-I, happened in 1968. By 1975, ENDF/B-IV had been released (in 1974). It is from that version that the earliest ENDF data were comprehensively made available to Los Alamos Monte Carlo code users. Over the next 45 years, data from all versions of ENDF have been processed and made available to MCNP® code users. As of the time of this article, the current ENDF version is ENDF/B-VIII.0.[88] The MCNP® code data libraries based on ENDF/B-VIII.0 are described in.[89]

Steady advances in accuracy, detail, and completeness in the ENDF evaluations have been made over the past 50 years. These advances have been the result of new experimental techniques, advanced nuclear physics models, and modern computing capabilities. In addition, testing of the data has become much more rigorous over time. All of this has led to much greater predictive capability for many more applications in codes such as the MCNP® code.

*IV.E.3. How Los Alamos Monte Carlo Codes Have Used Nuclear Data over the Years*

Documentation for the MCS code indicates a surprisingly robust collection of input options to describe nuclear data. Cross sections were allowed for total, absorption, elastic scattering, and inelastic scattering. Inelastic was further sub-divided into specific reaction channels, including fission. These data allowed for sampling of the distance to collision, the collision isotope, and the type of collision. Implicit capture was generally employed. Information to determine the outgoing neutron(s) energy and direction were also part of the code input.

For simple two-body reactions (elastic scattering or discrete level inelastic scattering), the scattering angle was sampled from distributions that were tabulated as a function of incident neutron energy. The distributions were either tabulated distributions or polynomial fits and could be specified in either the laboratory or center-of-mass system. The secondary energy was then determined from two-body kinematics with some approximations. For example, for elastic scattering on isotopes with a mass number greater than 25, the laboratory energy of the scattered neutron was assumed to be unchanged from the incident energy.

For more complex inelastic reactions, the scattering angle was sampled in the same manner as above. However, a host of data schemes were allowed for sampling secondary energies. There were tabulated distributions as a function of incident energy and tabulated distributions as a function of both incident energy and sampled scattering angle. There were formulas with requisite constants for an evaporation spectrum and a fission spectrum. There were a variety of possible specifications for (n,2n) reactions. One specification led to the creation of one secondary neutron with a doubling of weight. Other specifications enabled the first and second neutron to be sampled independently from different distributions – in these cases one neutron was banked. Finally, there was a formalism that was described as useful for some 3-body breakup reactions.

We have not found any examples of actual MCS input, so cannot say with what detail (for example in the number of incident energy (or velocity) points) all of these provided capabilities were actually used in practice.

The neutron thermal treatment in MCS was simply a thermal floor. Any neutron that was sampled with an energy below that floor was instead given an energy equal to the floor. The MCN code added a free-gas treatment for neutron thermalization with light isotopes.

Another nuclear-data feature discussed in MCN documentation is actually a capability implemented in the ENDF processing code (Probably ETOPL or MCPOINT at that time; predecessors of today's NJOY)[90] to both "thin" the number of energy points in the main cross-section grid after resonance reconstruction (to save storage) and conversely to add points where necessary (for accuracy when used with linear-linear interpolation). This brings up an important point, that it is not only the Monte Carlo code itself that determines how the original evaluated data are used, but also decisions made during processing (by the processing code or by user input to the processing code).

MCN also included an energy-dependent fission routine, with pre-fission neutrons emitted in an evaporation spectrum and the remaining neutrons in a fission spectrum. Both types of spectra depended upon the energy of the incoming neutron.

MCG was a companion code to MCN that performed Monte Carlo transport of gamma rays.[91] During the 1970's, MCN and MCG were combined in MCNG to enable coupled neutron and gamma ray transport.[92] This required neutron-induced gamma-production cross sections. The initial implementation of the gamma-production spectra was somewhat crude: "Currently the gamma production cross sections are used in multigroup form (30 neutron groups and 12 gamma groups), even though our transport codes use a continuous energy dependence on cross sections. This has been a matter of convenience." These multigroup data also allowed no separation of gammas produced by various reactions, for example fission gammas from capture gammas.

By 1980, the MCNP® code had emerged as the principal Monte Carlo product of Los Alamos. By that time, an S($\alpha,\beta$) capability had been added to the code as another neutron thermalization model. There remained issues associated with computer storage of the nuclear data – one result was the creation of so-called discrete reaction cross section libraries. These libraries binned the cross sections in 240 energy groups, but retained secondary distribution data in completely continuous energy format. Those discrete reaction became rather obsolete as computer memory expanded over the decades.

As mentioned previously, LA-UR-80-1158 asserted that "MCNP is a continuous-energy Monte Carlo code that makes no gross assumptions regarding data." While not questioning the veracity of that statement, it is worth noting that there have been many additional improvements to the MCNP® code since then regarding the types of data it utilizes and how the code uses those data. We briefly note some of those below (not necessarily complete nor in chronological order; references may be found on Ref. 93):

- Neutron-induced gamma production data were updated to be completely continuous in incident neutron and outgoing gamma energies. Additionally, the cross sections and spectra were separated to be specific to reactions as provided by ENDF data.
- In 1980, secondary angular distributions for non-isotropic particle-production were given, as a function of incident energy, by 32-equally probably cosine bins (again, a nod to memory limitations). That treatment was replaced by a continuous in cosine tabulation of scattering probabilities.
- MCNP® code users have always had the option of specifying either prompt or total fission nubar. However, earlier versions of the code used the same prompt fission energy data to sample all fission neutrons when the total nubar option was used. That approximation has been removed – now if a user specifies total nubar, the fraction of fission neutrons that are delayed are produced with a temporal and energy distribution as specified in the evaluation.
- Probability tables have been introduced to account for self-shielding in the unresolved energy region. These tables provide a distribution of cross sections at energies in the unresolved region, instead of simply the average cross section (which was used before probability tables were made available). The MCNP® code now samples the cross section when a neutron collision happens in the unresolved region.
- Early implementations of the S($\alpha,\beta$) capability provided secondary neutron data in either binned or discrete representations. A new treatment implements continuous distributions in secondary energy and angle.
- As ENDF representations of secondary data have become more complex and accurate, NJOY and the MCNP® code have been updated to handle those representations. In particular, a number of more detailed representations of correlated energy-angle distributions have been implemented.
- Neutron-induced charged particle cross sections and secondary spectra have been included in the data libraries and are used by the MCNP® code when

- specific charged particles are requested to be transported by the user. Decrements are made to neutron KERMA (heating) values in this case to allow the charged-particle energy to be deposited non-locally.
- Photonuclear data has been provided so that users can now couple neutrons and photons in both directions. Charged-particle incident data tables have also been made available.
- On-the-fly Doppler broadening capability has been included in the MCNP® code. This, plus a pre-processing capability in the MAKXSF code, enables users to utilize cross sections at temperatures specific to their needs and not be limited by the set of temperature-dependent data distributed with the code.
- A capability to use data for fission neutron multiplicity distributions has been added to the code. In nature, fission can result in 0, 1, 2, …. , 8 or more neutrons being produced. Evaluations (and therefore the MCNP® code) typically specify only the average nubar. There are applications that benefit from sampling nu from a multiplicity distribution.
- Event generator models such as CGMF and FREYA have been implemented in the MCNP® code. Such event generators use models that enable improved correlation of secondary particles in a way to conserve energy. Evaluated data, in general, provides no way for the MCNP® code to conserve energy or appropriately correlate secondary particles on a collision-by-collision basis, although the average results over many collisions are indeed faithful to the evaluated data.

While we feel justifiably proud of the rigor of the MCNP® code's nuclear data and physics capabilities, we are not ready to declare that there are no current approximations. We feel confident that additional advances will be made in the coming years.

### *IV.F. Los Alamos Monte Carlo MCNP® Code Today*

Large production software such as the MCNP® code have revolutionized science, not only in the way it is done, but also by becoming the repositories for physics knowledge. The MCNP® code represents over 750 person-years of sustained effort in code and data library development. The knowledge and expertise contained in the MCNP® code is formidable. Current MCNP® code development is characterized by a strong emphasis on quality control, documentation, and research. The MCNP® code undergoes rigorous testing using analytic, regression, and validation test sets to ensure correct operation.[94]

Los Alamos Monte Carlo codes and data libraries have continually improved from the ENIAC days to the present time to solve evermore complex problems. New features continue to be added to the code to reflect new advances in computer architectures, improvements in Monte Carlo methodology, and better physics models. Each year, about 1500 copies of the latest version of the MCNP® code are distributed to users all over the world by the Radiation Safety Information Computational Center. MCNP® classes are taught yearly on a wide range of topics. The goal of the Los Alamos Monte Carlo Group continues to be to provide the best Monte Carlo computational tool available to solve neutronics and other particle transport problems.

## V. 2021 LOS ALAMOS NEUTRONICS MONTE CARLO CALCULATIONS

The advances in Monte Carlo code capabilities since the ENIAC have mirrored the advances in computer technology. The predictive capability of the MCNP® code has increased enormously because of the physics and data improvements discussed previously. The MCNP® code can now be applied to wide range of problems from health physics, reactors, criticality safety, nuclear safeguards, high-energy physics, charged particle transport, shielding, design of complex experiments, and so on. The MCNP® user now has many controls about what physics, geometry types, variance reduction methods, data libraries, and tallies are to be used. The number of histories for a problem can exceed 1 billion. Graphics are available to visualize the Monte Carlo geometry, cross sections, and solutions.

## VI. SUMMARY

The Manhattan Project required the best theory and calculational capabilities possible to solve many problems, especially neutron diffusion. Unique conditions existed at Los Alamos during and after the Manhattan project to find these new solutions: momentous national security problems needed to be solved, brilliant scientists and engineers were assembled together with a common purpose, electronic computers were being developed, and adequate resources were made available. This resulted in a suite of neutron diffusion methods.

Los Alamos scientists created new deterministic - Sn - and stochastic - Monte Carlo - methods after the Manhattan Project to provide improved modeling for neutronics calculations. Both methods have been under continuous development at Los Alamos to provide the best computational tools possible for solving radiation transport problems. This paper has focused on the creation, development, implementation, and use of the Monte Carlo method at Los Alamos from those early days to the present.

Most of the required tasks to be able to make credible Monte Carlo calculations had never been done before. Consequently, there were many "firsts" that had to all be successfully completed: a new method had to be invented to model neutron diffusion, a recipe or algorithm needed to be written describing the technique, a programmable electronic digital computer needed to be designed and built, the computer had to be taught how to execute the steps in the algorithm, neutron cross sections had to be measured for selected materials, random numbers had to be available to perform the statistical sampling, and the computer program had to be debugged. The ENIAC Monte Carlo project accomplished all these tasks and produced successful calculations.

The Monte Carlo method and associated data libraries have been undergoing continuous development since the ENIAC experience. At Los Alamos, these developments have been encapsulated into the MCNP code. From 1963 and the first Los Alamos general purpose Monte Carlo code MCS to the present day MCNP, new and enhanced capabilities have been developed in physics, random number generators, variance reduction, geometries, tallies, graphics, programming techniques, multiprocessing, user interface, and verification and validation testing. MCNP and its predecessors have been run on every Los Alamos National Laboratory computer. MCNP is applied to increasingly difficult problems because of its well-known predictive capability. The feedback from our users worldwide is critical to keeping MCNP as a highly-respected and reliable computational tool for solving radiation transport problems.

The ENIAC Monte Carlo calculations provided the proof-of-principle of the viability of the Monte Carlo method being used on computers. Today, Monte Carlo is used virtually everywhere when difficult problems involving random processes need to be solved. It all started with the ENIAC experience.